\documentclass[]{spie}  

\usepackage{amsmath,amsfonts,amssymb}
\usepackage{graphicx}
\usepackage{caption}
\usepackage{subcaption}
\usepackage[colorlinks=true, allcolors=blue]{hyperref}

\title{Application Checkpoint and Power Study on Large Scale Systems}

\author[$\dag$]{Yuping Fan}
\affil[$\dag$]{Illinois Institute of Technology} 

\begin{document} 
\maketitle

\begin{abstract}
Power efficiency is critical in high performance computing (HPC) systems. To achieve high power efficiency on application level, it is vital importance to efficiently distribute power used by application checkpoints. In this study, we analyze the relation of application checkpoints and their power consumption. The observations could guide the design of power management.
\end{abstract}

\keywords{High Performance Computing (HPC), power management, performance analysis}

\section{INTRODUCTION}
\label{sec:intro}  

As the size of HPC systems rapidly increasing, the amount of electrical power consumed by HPC systems are keeping increasing. As a result, power becomes the leading constraint for the design of the next generation HPC systems. There are several existing solutions, such as power capping and dynamic voltage and frequency scaling (DVFS), to address the power constraints. To efficient distribute power, it is crucial to analyze application behaviors and provide customized power policies for various applications.

Checkpointing is the widely used fault tolerate mechanism. Checkpoints can be categorized into system-level and application-level checkpoints. Application-level checkpoints aim to recover failed applications. Applications can choose their own checkpoint frequency and other parameters related to checkpoints. 

Power and time consumed by checkpointing is not negligible. Therefore, this work studies how the checkpoint and faults affect the power and energy consumption. We utilize the MonEQ \cite{MonEQ} to gather the power information on Mira. MonEQ is a user-level profiling library, which collects power information at node card level. Beside the node card power information as a whole, the power information breaks down into six domains, which are dram voltage, link chip voltage, SRAM voltage, optics voltage, optics voltage, PCIExpress voltage and link chip core voltage. Three benchmarks, NPB \cite{NPB}, Flash \cite{flash} and Stream \cite{stream}, are investigated in this study. For NPB and Stream benchmarks, we add the checkpoints and faults by FTI \cite{FTI}. FTI is a fault tolerance interface, which provides application level checkpointing for large-scale supercomputers. Four configurable checkpoint levels are offered in FTI, which provides different levels of protection for applications. The checkpoint frequencies can be configured and be optimized \cite{Di} to achieve the balance between execution time and program correctness. Flash benchmark has its own checkpointing strategy, but there is no option to inject faults. 

The observations in this study can guide HPC job scheduler to intelligently schedule jobs \cite{Fan1, Qiao1, Fan2, Li1, Fan3, Topper, Fan4, Qiao2, Fan5, Fan6, Fan7} and wisely set checkpoint strategies in order to achieve high power efficiency.

The remainder of this paper is organized as follows. We start by
introducing background and related work in \S\ref{sec:background}. \S\ref{sec:analysis} presents the analysis and observation. We conclude the paper in \S\ref{sec:conclusion}.

\section{Background}\label{sec:background}

This section introduces HPC power management in \S\ref{sec:HPC power management} and Application Checkpointing \S\ref{sec:Application Checkpointing}. Then, we introduce the benchmarks analyzed in this study (\S\ref{sec:benchmarks}).

\subsection{HPC Power Management}
\label{sec:HPC power management}

As the size of HPC systems size rapidly grow, the limited power budget becomes one of the most crucial challenges. Several hardward power management mechanisms, such as dynamic voltage and frequency scaling (DVFS) and power capping, have been developed. Power can be management either at the system level or at the application level. Studies show that managing power at application level is more efficient. This is because the effects of power capping on different applications vary. Even the effects of power capping on different stages of an applications vary. Therefore, to efficient manage power, it is crucial to comprehensively analyze the effects of power management on individual applications.

\subsection{Application Checkpointing}
\label{sec:Application Checkpointing}
Fault tolerance is a serious problem in HPC systems. HPC systems consist of  millions of components. A single point of failure could lead to application failure, or even system failure. As the HPC system sizes rapidly increasing, the failures are happened more frequently.  Checkpointing is one of the most widely used fault tolerance techniques. HPC checkpointing can be implemented in two levels: system level and application level. System level checkpointing can prevent catastrophic system failures, but it is very time- and memory- intensive. On the other hand, application checkpointing is more lightweight and can prevent application failures. Typically, application checkpointing is scheduled more often than system checkpointing.

\subsection{Benchmarks}\label{sec:benchmarks}
In order to comprehensively evaluate various types of applications performance, three benchmarks are studied in this work.

\subsubsection{NPB benchmarks}
The NAS Parallel Benchmarks (NPB) are a small set of programs designed to help evaluate the performance of highly parallel supercomputers. The benchmarks are derived from computational fluid dynamics (CFD) applications and consist of five kernels and three pseudo-applications. We study three representative programs from NPB benchmarks:
\begin{enumerate}
    \item \textbf{CG}: Conjugate Gradient, irregular memory access and communication
    \item \textbf{FT}: discrete 3D fast Fourier Transform, all-to-all communication
    \item \textbf{LU}: Lower-Upper Gauss-Seidel solver
\end{enumerate}

\subsubsection{Flash}

Flash benchmark is multiphysics multiscale simulation code. We select Sedov explosion problem in this study. The Sedov explosion problem is a hydrodynamical test to check the code's ability to deal with strong shocks and non-planar symmetry. 

\subsubsection{STREAM}

STREAM is a simple, synthetic benchmark designed to measure sustainable memory bandwidth (in MB/s) and a corresponding computation rate for four simple vector kernels. STREAM is the representative of memory intensive jobs. 

\section{Analysis}\label{sec:analysis}
\subsection{NPB benchmarks}
For NPB benchmarks, we choose the largest problem size E to do all the experiments. 

In order to show the fault tolerance influence on power consumption, each application ran three times with the same settings, except the checkpoints and faults. The first run records the power information without checkpoints and faults. The second run adds checkpoints but no injected faults. The third run adds both checkpoints and injected faults.

\begin{figure}
    \centering
     \begin{subfigure}[b]{0.45\textwidth}
         \centering
         \includegraphics[width=\textwidth]{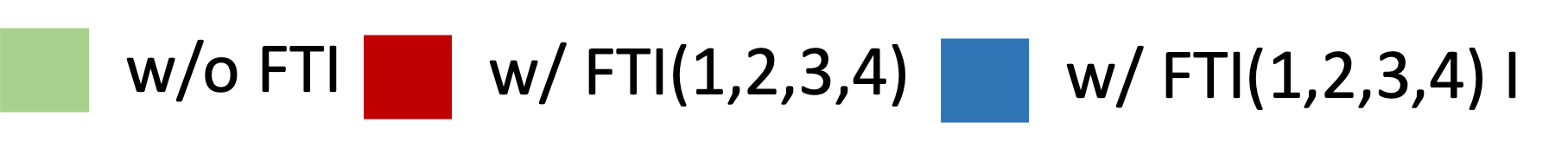}
     \end{subfigure}
     \\
     \centering
     \begin{subfigure}[b]{0.32\textwidth}
         \centering
         \includegraphics[width=\textwidth]{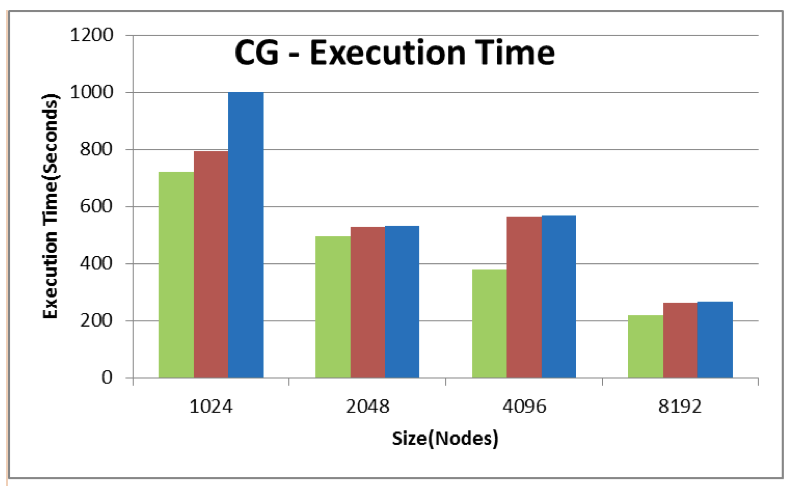}
     \end{subfigure}
     \hfill
     \begin{subfigure}[b]{0.32\textwidth}
         \centering
         \includegraphics[width=\textwidth]{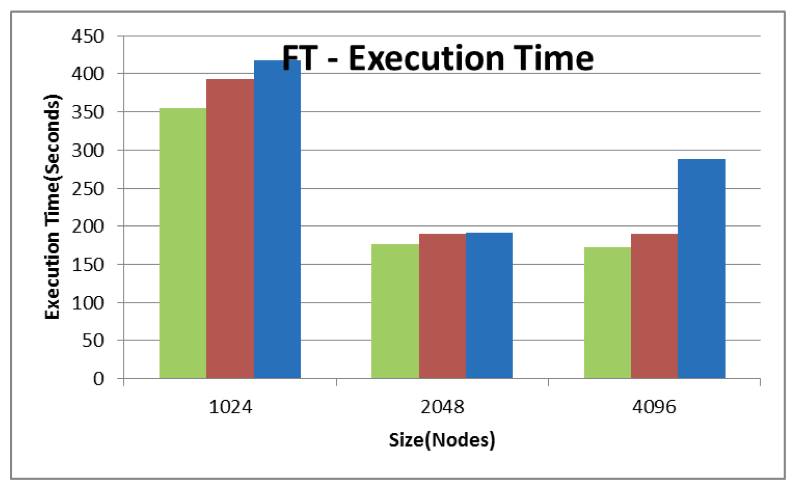}
     \end{subfigure}
     \hfill
     \begin{subfigure}[b]{0.32\textwidth}
         \centering
         \includegraphics[width=\textwidth]{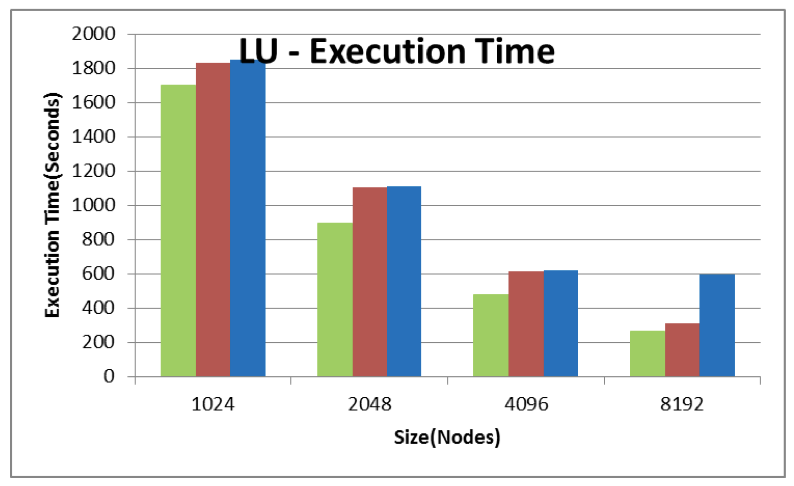}
     \end{subfigure}
        \caption{The execution time of NPB benchmarks.The green bars represents the execution time without checkpoints; the red bars represents the execution time with checkpoints but without fault injections; the blue bars represents the execution time with checkpoints and fault injections.}
        \label{NPB_execution_time_barplot}
\end{figure}

\begin{figure}
    \centering
     \begin{subfigure}[b]{0.45\textwidth}
         \centering
         \includegraphics[width=\textwidth]{legend.png}
     \end{subfigure}
     \\
     \centering
     \begin{subfigure}[b]{0.32\textwidth}
         \centering
         \includegraphics[width=\textwidth]{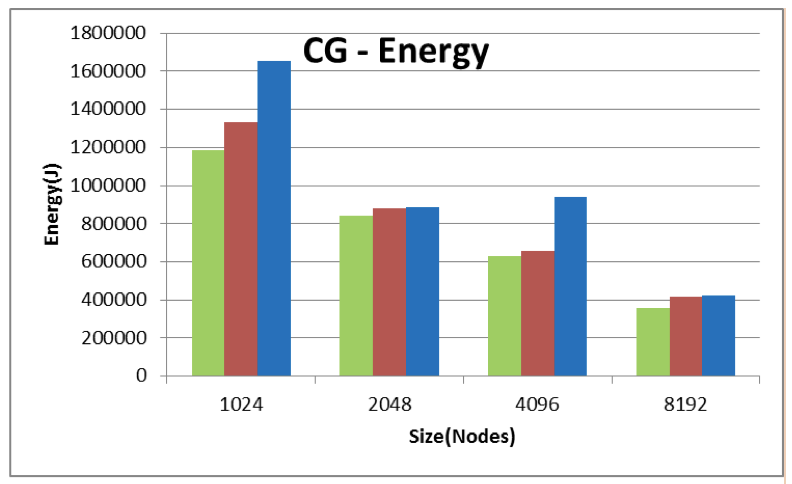}
     \end{subfigure}
     \hfill
     \begin{subfigure}[b]{0.32\textwidth}
         \centering
         \includegraphics[width=\textwidth]{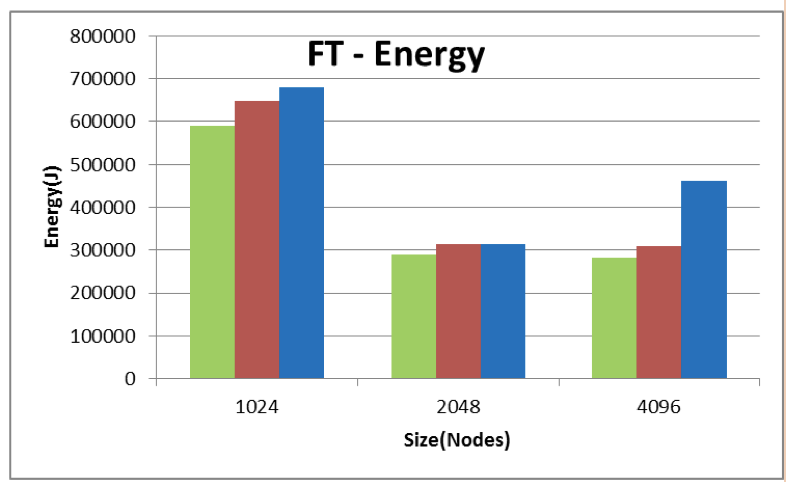}
     \end{subfigure}
     \hfill
     \begin{subfigure}[b]{0.32\textwidth}
         \centering
         \includegraphics[width=\textwidth]{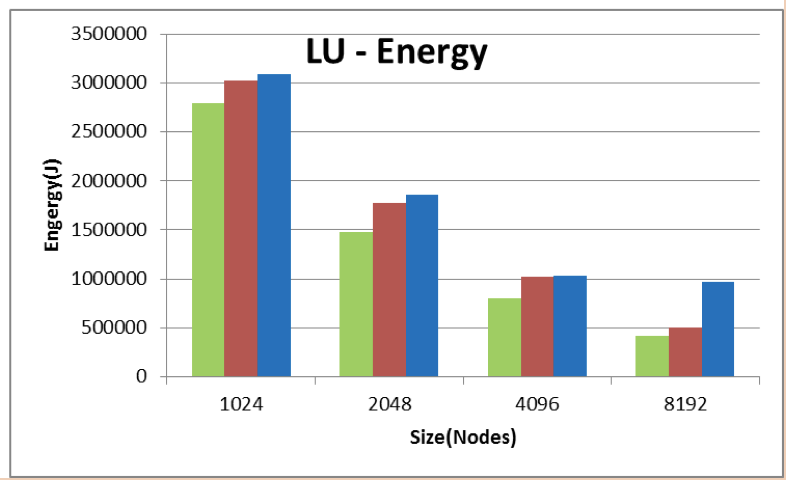}
     \end{subfigure}
        \caption{The energy consumption of NPB benchmarks.}
        \label{NPB_energy_barplot}
\end{figure}

Figure \ref{NPB_execution_time_barplot} and figure \ref{NPB_energy_barplot} presents the effects of checkpoints on execution time and energy consumption respectively.

\begin{table}[]
\caption{Incremental percentage of average power, execution time and energy. Negative percentage denotes the decrease, for example, the power consumption with FTI of the NPB: CG on 2048 nodes decrease 1.71\% compare to that without FTI.} 
\label{tab:NPB_table}
\resizebox{\textwidth}{!}{
\begin{tabular}{|l|l|l|l|l|l|l|l|l|l|}
\hline
     & \multicolumn{3}{l|}{with FTI/without FTI} & \multicolumn{3}{l|}{Inject Fault/without FTI} & \multicolumn{3}{l|}{Inject Fault/with FTI} \\ \hline
CG   & Average Power  & Execution Time & Energy  & Average Power   & Execution Time  & Energy    & Average Power  & Execution Time  & Energy  \\ \hline
1024 & 0.98\%         & 10.00\%        & 12.43\% & 0.30\%          & 38.34\%         & 39.61\%   & -0.68\%        & 25.77\%         & 24.17\% \\ \hline
2048 & -1.71\%        & 6.41\%         & 4.59\%  & -1.72\%         & 6.75\%          & 4.92\%    & 0.61\%         & 0.32\%          & 0.31\%  \\ \hline
4096 & 0.04\%         & 38.85\%        & 4.17\%  & -0.25\%         & 47.79\%         & 49.06\%   & 0.29\%         & 0.64\%          & 12.60\% \\ \hline
8192 & -3.58\%        & 19.79\%        & 15.50\% & -2.50\%         & 20.79\%         & 17.77\%   & 1.12\%         & 0.84\%          & 1.97\%  \\ \hline
\multicolumn{10}{|l|}{FT}                                                                                                                     \\ \hline
1024 & -0.48\%        & 10.41\%        & 9.87\%  & -1.87\%         & 17.33\%         & 15.14\%   & -1.39\%        & 6.27\%          & 4.79\%  \\ \hline
2048 & 0.25\%         & 7.94\%         & 8.21\%  & 0.25\%          & 8.27\%          & 8.55\%    & 0.01\%         & 0.30\%          & 0.31\%  \\ \hline
4096 & -0.92\%        & 10.77\%        & 9.75\%  & -2.00\%         & 67.49\%         & 64.14\%   & -1.09\%        & 51.21\%         & 49.57\% \\ \hline
\multicolumn{10}{|l|}{LU}                                                                                                                     \\ \hline
1024 & 0.96\%         & 7.49\%         & 8.52\%  & 1.83\%          & 8.72\%          & 10.71\%   & 0.86\%         & 1.15\%          & 2.02\%  \\ \hline
2048 & -2.38\%        & 22.93\%        & 20.00\% & 1.28\%          & 23.81\%         & 25.40\%   & 3.76\%         & 0.71\%          & 4.49\%  \\ \hline
4096 & -0.42\%        & 28.48\%        & 27.95\% & -0.38\%         & 29.58\%         & 29.09\%   & 0.04\%         & 0.85\%          & 0.90\%  \\ \hline
8192 & -0.24\%        & 16.31\%        & 19.72\% & 1.52\%          & 121.78\%        & 129.38\%  & -0.07\%        & 90.69\%         & 91.60\% \\ \hline
\end{tabular}
}
\end{table}

The following subsection gives more detailed analysis on these three benchmarks.

\subsubsection{CG}
\begin{figure}
     \centering
     \begin{subfigure}[b]{0.45\textwidth}
         \centering
         \includegraphics[width=\textwidth]{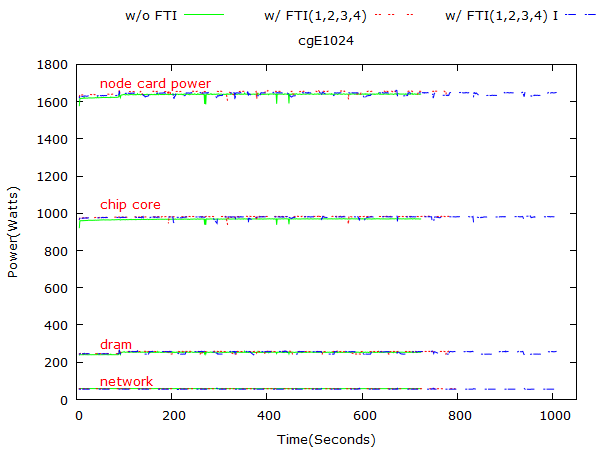}
     \end{subfigure}
     \hfill
     \begin{subfigure}[b]{0.45\textwidth}
         \centering
         \includegraphics[width=\textwidth]{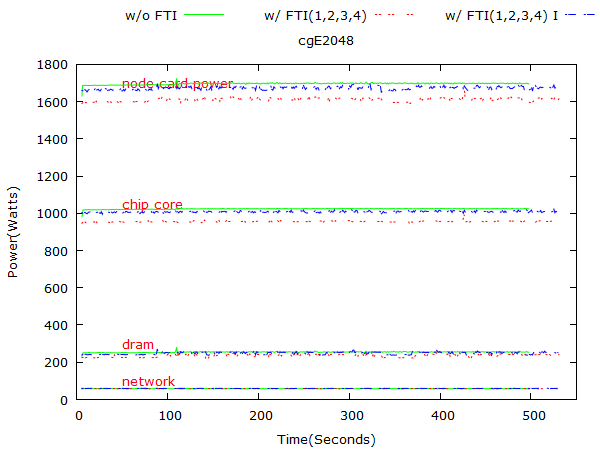}
     \end{subfigure}
     \hfill
     \begin{subfigure}[b]{0.45\textwidth}
         \centering
         \includegraphics[width=\textwidth]{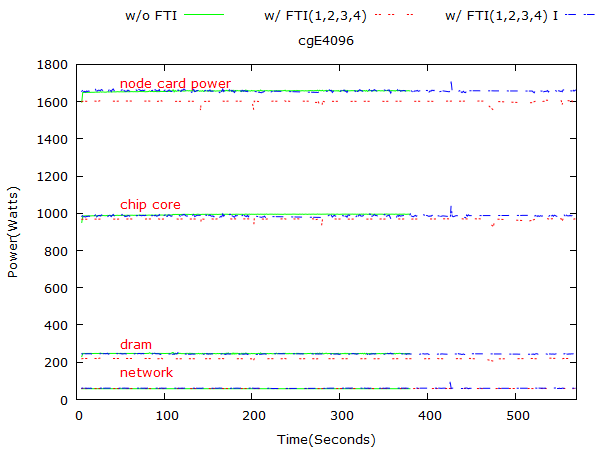}
     \end{subfigure}
     \hfill
     \begin{subfigure}[b]{0.45\textwidth}
         \centering
         \includegraphics[width=\textwidth]{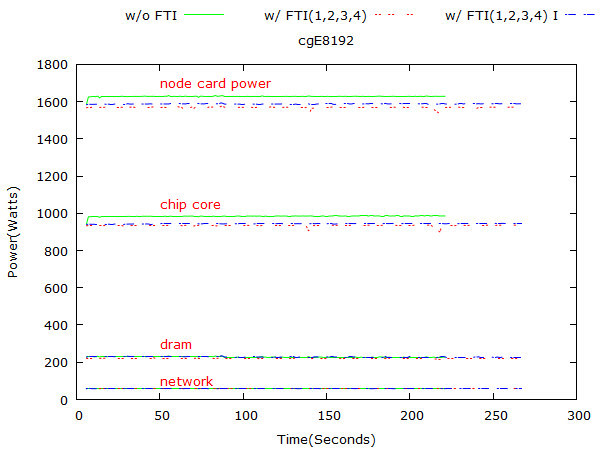}
     \end{subfigure}
        \caption{NPB: CG power consumption on 1024, 2048, 4096 and 8192 nodes respectively. “w/o FTI” denotes experiment without checkpoints and faults. “w/ FTI(1,2,3,4)” denotes experiment with FTI and the checkpointing frequencies from level 1 to level 4 are 1, 2, 3 and 4 respectively. “w/ FTI(1,2,3,4) I” denotes the experiment with FTI and inject faults.}
        \label{cg_plot}
\end{figure}

\begin{figure}
     \centering
     \begin{subfigure}[b]{0.45\textwidth}
         \centering
         \includegraphics[width=\textwidth]{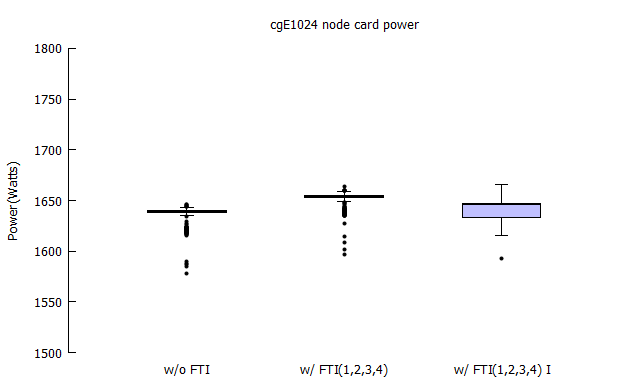}
     \end{subfigure}
     \hfill
     \begin{subfigure}[b]{0.45\textwidth}
         \centering
         \includegraphics[width=\textwidth]{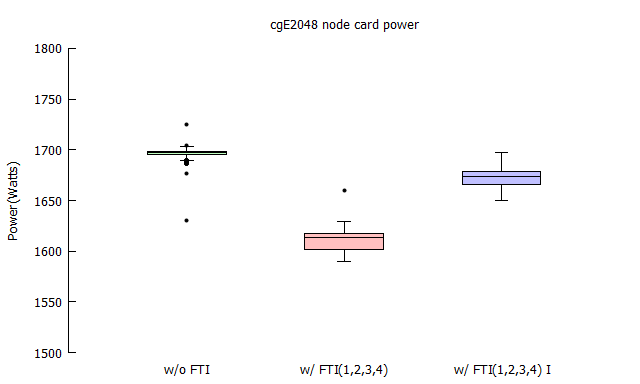}
     \end{subfigure}
     \hfill
     \begin{subfigure}[b]{0.45\textwidth}
         \centering
         \includegraphics[width=\textwidth]{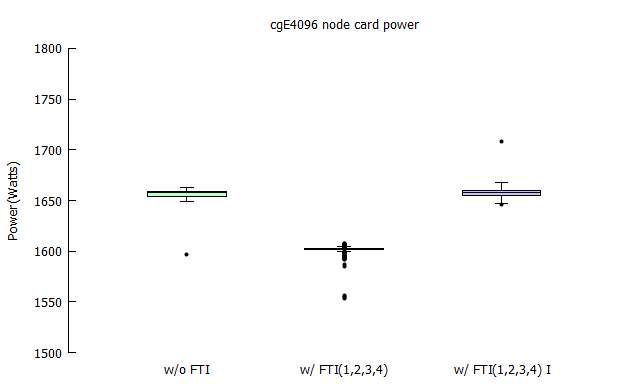}
     \end{subfigure}
     \hfill
     \begin{subfigure}[b]{0.45\textwidth}
         \centering
         \includegraphics[width=\textwidth]{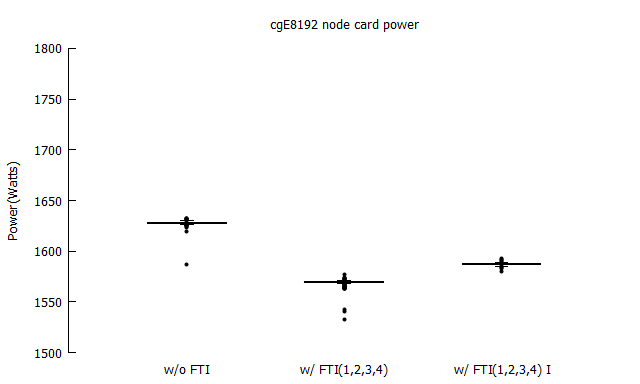}
     \end{subfigure}
        \caption{Box plot comparison of average power consumption at the node card level for NPB: CG.}
        \label{cg_boxplot}
\end{figure}

Figure \ref{cg_plot} and figure \ref{cg_boxplot} show the CG program power consumption on 1024, 2048, 4096 and 8192 nodes respectively. CG programs ran at four scales. They are 1k, 2k, 4k and 8k nodes, which means 1k, 2k, 4k and 8k MPI ranks separately. Since chip core, dram, networks consume most energy and checkpoints and faults do not significant influences on other domains, we ignore other domains. 

\subsubsection{FT}
\begin{figure}
     \centering
     \begin{subfigure}[b]{0.32\textwidth}
         \centering
         \includegraphics[width=\textwidth]{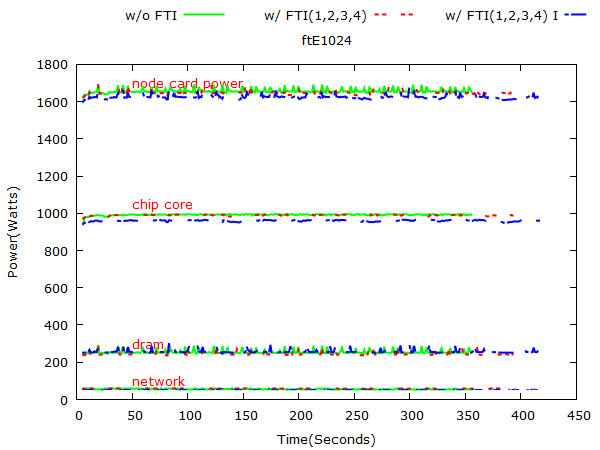}
     \end{subfigure}
     \hfill
     \begin{subfigure}[b]{0.32\textwidth}
         \centering
         \includegraphics[width=\textwidth]{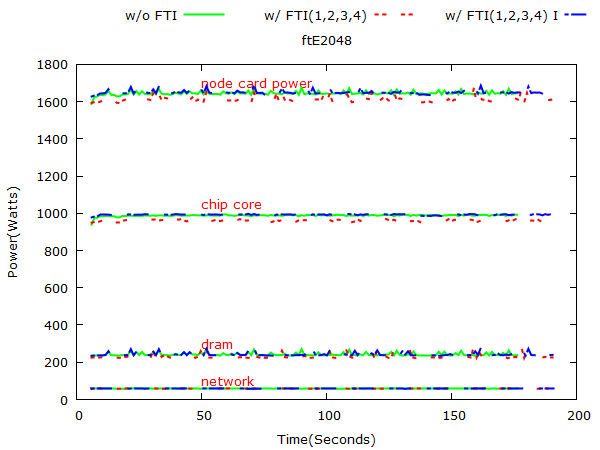}
     \end{subfigure}
     \hfill
     \begin{subfigure}[b]{0.32\textwidth}
         \centering
         \includegraphics[width=\textwidth]{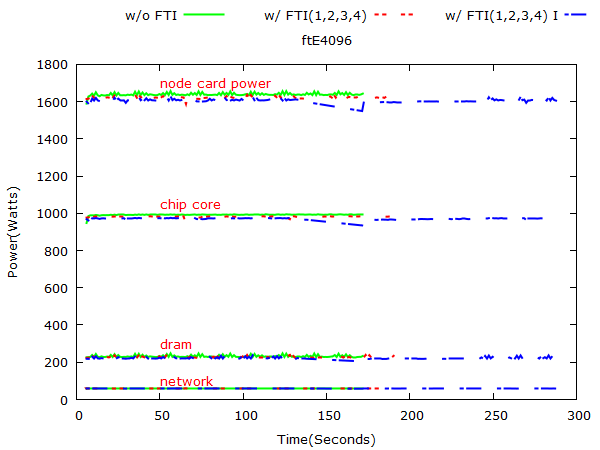}
     \end{subfigure}
        \caption{NPB: FT power consumption on 1024, 2048 and 4096 nodes respectively.}
        \label{ft_plot}
\end{figure}

\begin{figure}
     \centering
     \begin{subfigure}[b]{0.32\textwidth}
         \centering
         \includegraphics[width=\textwidth]{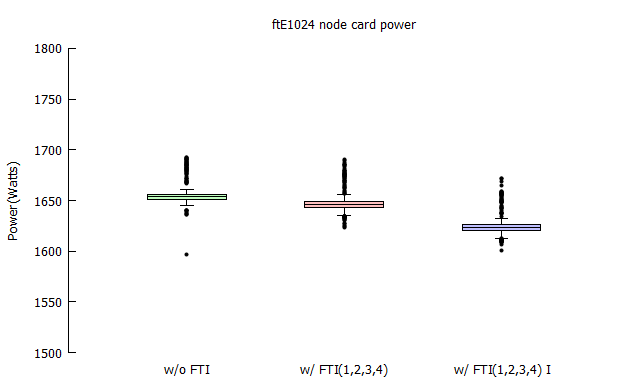}
     \end{subfigure}
     \hfill
     \begin{subfigure}[b]{0.32\textwidth}
         \centering
         \includegraphics[width=\textwidth]{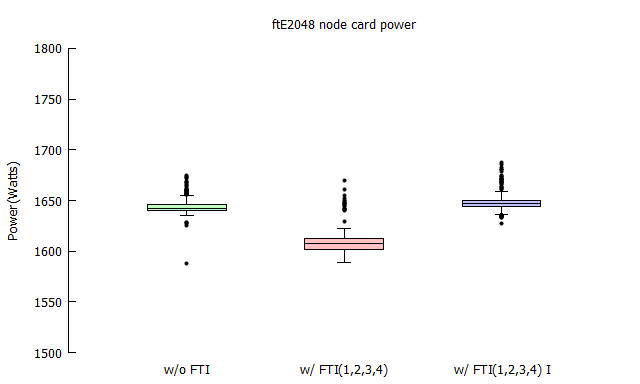}
     \end{subfigure}
     \hfill
     \begin{subfigure}[b]{0.32\textwidth}
         \centering
         \includegraphics[width=\textwidth]{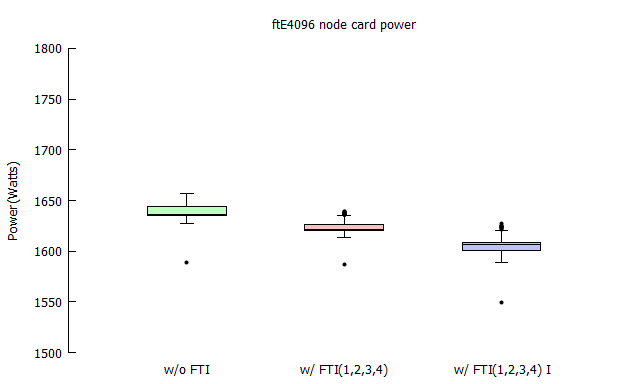}
     \end{subfigure}
        \caption{Box plot comparison of average power consumption at the node card level for NPB: FT.}
        \label{ft_boxplot}
\end{figure}

Figure \ref{ft_plot} and \ref{ft_boxplot} show FT's power consumption on different problem sizes. The FT experiment ran on 1k, 2k, and 4k nodes. We skip the experiment on 8k nodes, because FT experiment on 8k nodes is too short to gather enough power consumption information. 

\subsubsection{LU}
\begin{figure}
     \centering
     \begin{subfigure}[b]{0.45\textwidth}
         \centering
         \includegraphics[width=\textwidth]{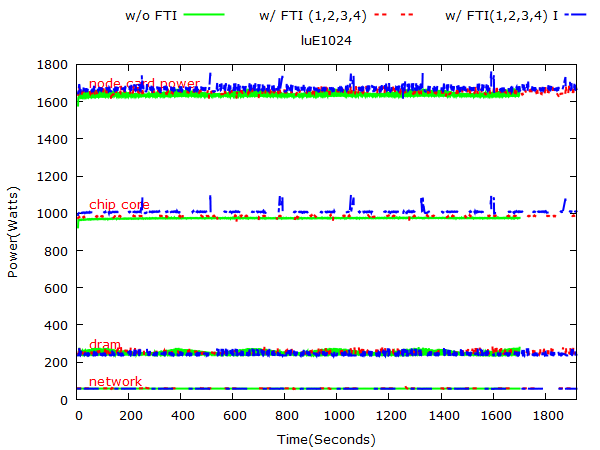}
     \end{subfigure}
     \hfill
     \begin{subfigure}[b]{0.45\textwidth}
         \centering
         \includegraphics[width=\textwidth]{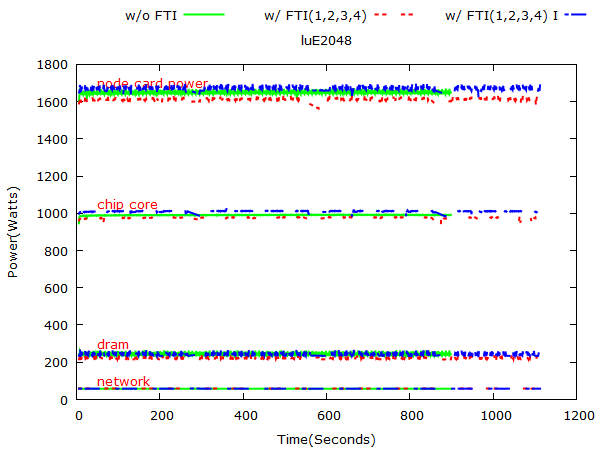}
     \end{subfigure}
     \hfill
     \begin{subfigure}[b]{0.45\textwidth}
         \centering
         \includegraphics[width=\textwidth]{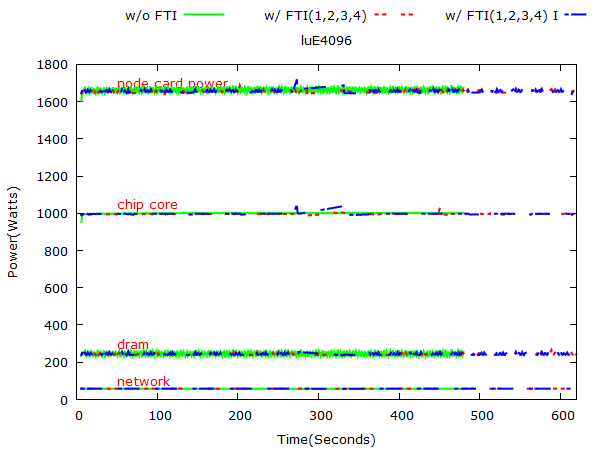}
     \end{subfigure}
     \hfill
     \begin{subfigure}[b]{0.45\textwidth}
         \centering
         \includegraphics[width=\textwidth]{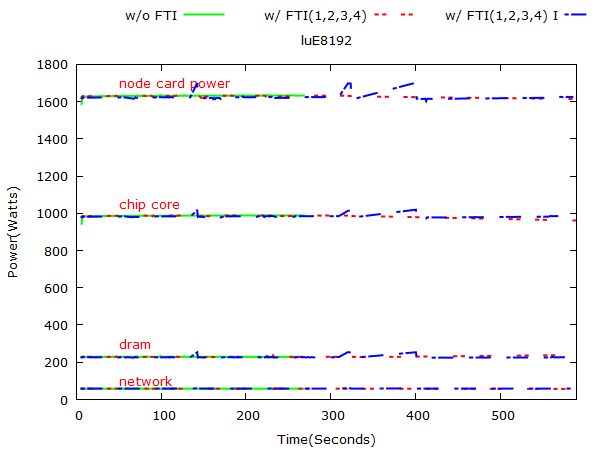}
     \end{subfigure}
        \caption{NPB: LU power consumption on 1024, 2048, 4096 and 8192 nodes respectively.}
        \label{lu_plot}
\end{figure}

\begin{figure}
     \centering
     \begin{subfigure}[b]{0.45\textwidth}
         \centering
         \includegraphics[width=\textwidth]{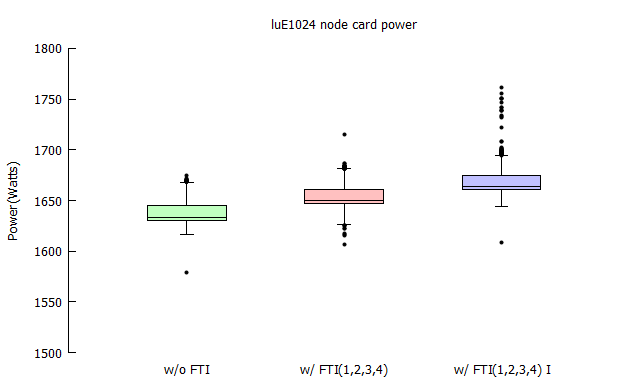}
     \end{subfigure}
     \hfill
     \begin{subfigure}[b]{0.45\textwidth}
         \centering
         \includegraphics[width=\textwidth]{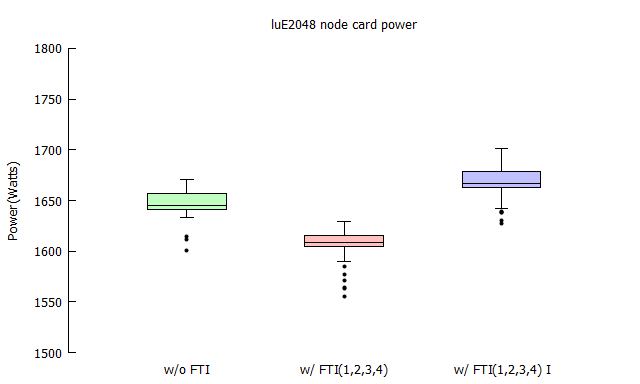}
     \end{subfigure}
     \hfill
     \begin{subfigure}[b]{0.45\textwidth}
         \centering
         \includegraphics[width=\textwidth]{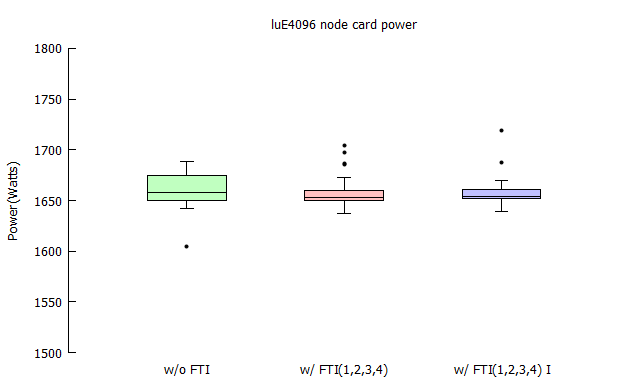}
     \end{subfigure}
     \hfill
     \begin{subfigure}[b]{0.45\textwidth}
         \centering
         \includegraphics[width=\textwidth]{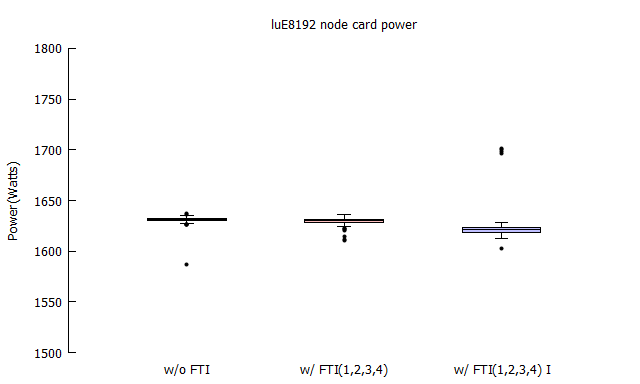}
     \end{subfigure}
        \caption{Box plot comparison of average power consumption at the node card level for NPB: LU.}
        \label{lu_boxplot}
\end{figure}

Figure \ref{lu_plot} and \ref{lu_boxplot} show LU's power consumption on different problem sizes.

\subsection{Flash}
In-build checkpoint option in flash code enables us doing checkpoint at regular time or instruction intervals. In this experiment, we compare the power consumption of Sedov ran on 512 nodes with checkpoints and that with checkpoints.

\begin{figure}
     \centering
    \includegraphics[width=\textwidth]{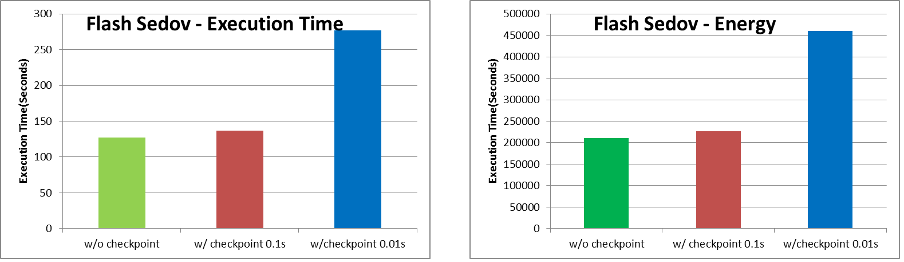}
    \caption{The execution time and energy of Flash: Sedov application.}
    \label{flash_barplot}
\end{figure}

Figure \ref{flash_barplot} presents the execution time and energy consumption of Flash: Sedov application.

\begin{figure}
     \centering
     \begin{subfigure}[b]{0.45\textwidth}
         \centering
         \includegraphics[width=\textwidth]{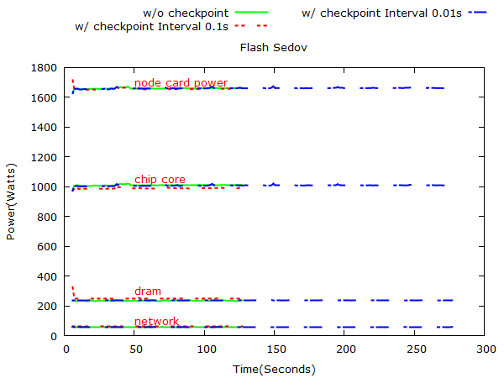}
     \end{subfigure}
     \hfill
     \begin{subfigure}[b]{0.45\textwidth}
         \centering
         \includegraphics[width=\textwidth]{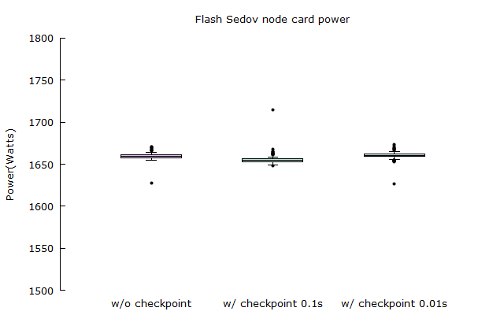}
     \end{subfigure}
        \caption{Flash: Sedov power consumption at 512 scale.}
        \label{flash_plot}
\end{figure}

Figure \ref{flash_plot} represents Flash's power consumption at 512 scale.

\subsection{STREAM}
\begin{figure}
     \centering
    \includegraphics[width=\textwidth]{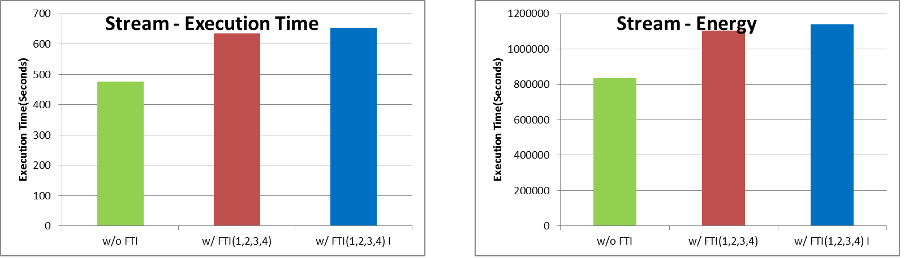}
    \caption{The execution time and energy of STREAM benchmark.}
    \label{stream_barplot}
\end{figure}

Note that for STREAM, FTI is used to do regular checkpoints and random fault injection.
Figure \ref{stream_barplot} presents the execution time and energy consumption of STREAM.

\begin{figure}
     \centering
     \begin{subfigure}[b]{0.45\textwidth}
         \centering
         \includegraphics[width=\textwidth]{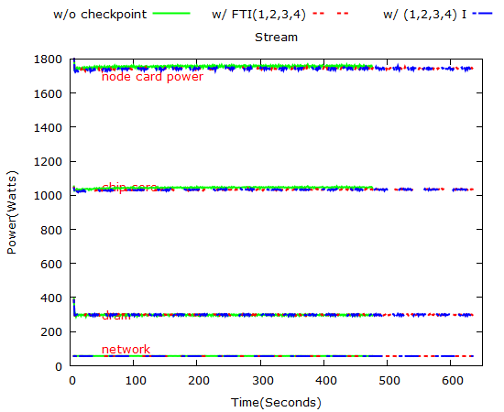}
     \end{subfigure}
     \hfill
     \begin{subfigure}[b]{0.45\textwidth}
         \centering
         \includegraphics[width=\textwidth]{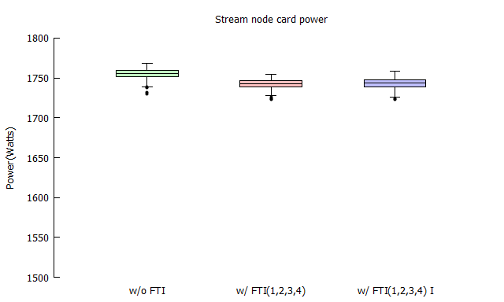}
     \end{subfigure}
        \caption{STREAM power consumption on 32 nodes.}
        \label{stream_plot}
\end{figure}

Figure \ref{stream_plot} represents STREAM's power consumption on 32 nodes.

\subsection{Key Observations}
Based on the comprehensive analysis on NPB, Flash, and Stream benchmarks, we summarize the key observations with the follwing five categories.

\subsubsection{With checkpoint or without checkpoint}
\begin{itemize}
    \item The power consumption remains the same or reduces a little(-3.58\%~-0.24\%) in most cases by adding checkpoint without injecting faults. The possible reason is that checkpointing is not computation intensive task, hence the chip core power is reduced in most cases, which causes the total power reduced too. There are only four exceptions, which are luE1024, luE2048, ftE2048 and cgE1024, but the differences are trivial and have not exceeded 1\% in all three cases.  
    \item The execution time and energy increase after adding checkpoints. Adding the checkpoints with the same frequencies have the different effect on different application and even the same application running on different scales. For NPB benchmarks, the increases in the time range from 6.41\% to 38.85\% and the increases in the energy range from 4.17\% to 27.95\%. We can see from figure 7, that the energy cost is close related to execution time and they follow the same trend. For Stream benchmark, the execution time increase 33.15\% and the energy increases 31.99\% by adding checkpoints. Stream benchmark is more sensitive to the checkpoints. This is because Stream is a memory-intensive application. There is no local disk on nodes, FTI saves the level 1, 2 and 3 checkpoints in memory, which cause the competition between the checkpoint and progress of stream and we can observe the higher DRAM power than other benchmarks in this study. Hence, we observe more significant influence on Stream benchmark comparing with NPB benchmarks.
\end{itemize}

\subsubsection{Inject fault or not}
\begin{itemize}
    \item Injecting faults cause the power increases or remain the same(-0.68\%~3.76\%) in most cases comparing to only adding checkpoints without faults. The main increases came from the chip core power increases because the recovery process is computationally intensive. From Table \ref{tab:NPB_table}, we observe that CG and LU average power on all 4 different scales increase the power consumption. Adding faults seems to reduce the power consumption for FT, this is because FT benchmark runs too short, that these two experiments do a different number of the four-level checkpoint and the difference of the number of checkpoints have a great influence on the short jobs. Stream benchmark and Flash benchmark also show the same trend.
    \item Another finding is that injecting faults make the power consumption fluctuate. We can see it from the boxplots. It is clear that there are more outliers for power consumption with fault injections. There are more spikes after injecting faults. The recovery process first retrieves information from disk then re-computes from the latest checkpoint. The retrieving process is the not computationally intensive, but the re-computation is. The interruption of faults explains why there are more ups and downs when injecting faults.
    \item Randomly injected faults have different effects on execution time and energy. Figure 7 shows that injected faults make the execution time of cgE1024, ftE4096 and luE8192 increase significantly. The logs of these runs show that all these runs experienced the fault error and recovered from level 4 checkpoint, which is the most time-consuming recovery process. Another observation is that as the number of nodes increases, the recovery process takes more time. For example, a fault error of luE8192 takes 90.69\% more than that without error. 
    \item If we ignore the cases that experienced a fatal error, FTI is very efficient in recovery. The incremental percentage of execution time is ranging from 0.30\% to 6.27\% and that of energy is ranging from 0.31\% to 4.79\%. In reality, the failure rate is lower than the rate in our experiment, because we injected faults. Hence, we can say that if there are no errors that requiring all the nodes back to last checkpoints, faults have a trivial effect on execution time and energy. Stream also shows the same feature that injects faults without a fatal error does not affect the execution time and energy too much.
\end{itemize}

\subsubsection{Number of nodes}
\begin{itemize}
    \item For CG and FT, increasing the number of nodes seems further lowered the power with checkpointing comparing to that without checkpointing. The difference may be the result of the communication delay by introducing checkpoint. Level 2 and level 3 require storing checkpoints at other nodes, which involves the network. More nodes mean there are more communication and information needed to transfer in the system, which may delay the computation. As the result, the chip core power goes down as the number of node increases. 
\end{itemize}

\subsubsection{Energy}
\begin{itemize}
    \item From the energy figures and Table \ref{tab:NPB_table}, we can see that running application without checkpoints consumes less energy than running the same application with checkpoints. It is also true that application with fault injection consumes more energy than application without fault injection. 
\end{itemize}

\subsubsection{Time}
\begin{itemize}
    \item The column plots present that running application without checkpoint takes less time and running application with fault takes more time than that without faults.
    \item The time has a dominate role in energy. Although larger scale results in shorter execution time, the total energy consumed by application increase as the scale becomes larger. From energy usage view, smaller scale help saves energy.
\end{itemize}

\section{Conclusion}\label{sec:conclusion}
In this paper, we have analyze the effect of application checkpoints on power consumption. The observations provides insight to design power- and checkpoint-aware scheduling policies.

 

\bibliography{report.bib} 

\begin{thebibliography}{10}

\bibitem{MonEQ}
Wallace, S., Z.~Zhou, V.~V., Coghlan, S., Tramm, J., Lan, Z., and Papka, M.~E.,
  ``Application power profiling on ibm blue gene/q,'' in [{\em IEEE
  International Conference on Cluster Computing
  (CLUSTER)}{\nolinebreak\hspace{0.1em}]},  (2013).

\bibitem{NPB}
``{NAS Parallel Benchmarks}.''
  \url{https://www.nas.nasa.gov/software/npb.html}.
\newblock (Accessed: 27 August 2021).

\bibitem{flash}
``{Flash}.'' \url{http://flash.uchicago.edu/site/}.
\newblock (Accessed: 27 August 2021).

\bibitem{stream}
``{Stream}.'' \url{https://www.cs.virginia.edu/stream/}.
\newblock (Accessed: 27 August 2021).

\bibitem{FTI}
``{Fault Tolerance Interface (FTI)}.'' \url{http://leobago.com/projects/}.
\newblock (Accessed: 27 August 2021).

\bibitem{Di}
Di, S., Bautista-Gomez, L., and Cappello, F., ``Optimization of multi-level
  checkpoint model with uncertain execution scales,'' in [{\em Proceedings of
  the International Conference for High Performance Computing, Networking,
  Storage and Analysis (SC)}{\nolinebreak\hspace{0.1em}]},  (2014).

\bibitem{Fan1}
Fan, Y., Rich, P., Allcock, W., Papka, M., and Lan, Z., ``{Trade-Off Between
  Prediction Accuracy and Underestimation Rate in Job Runtime Estimates},'' in
  [{\em CLUSTER}{\nolinebreak\hspace{0.1em}]},  (2017).

\bibitem{Qiao1}
Qiao, P., Wang, X., Yang, X., Fan, Y., and Lan, Z., ``{Preliminary Interference
  Study About Job Placement and Routing Algorithms in the Fat-Tree Topology for
  HPC Applications},'' in [{\em CLUSTER}{\nolinebreak\hspace{0.1em}]},  (2017).

\bibitem{Fan2}
Allcock, W., Rich, P., Fan, Y., and Lan, Z., ``{Experience and Practice of
  Batch Scheduling on Leadership Supercomputers at Argonne},'' in [{\em
  JSSPP}{\nolinebreak\hspace{0.1em}]},  (2017).

\bibitem{Li1}
Li, B., Chunduri, S., Harms, K., Fan, Y., and Lan, Z., ``{The Effect of System
  Utilization on Application Performance Variability},'' in [{\em
  ROSS}{\nolinebreak\hspace{0.1em}]},  (2019).

\bibitem{Fan3}
Fan, Y., Lan, Z., Rich, P., Allcock, W., Papka, M., Austin, B., and Paul, D.,
  ``{Scheduling Beyond CPUs for HPC},'' in [{\em
  HPDC}{\nolinebreak\hspace{0.1em}]},  (2019).

\bibitem{Topper}
Yu, L., Zhou, Z., Fan, Y., Papka, M., and Lan, Z., ``{System-wide Trade-off
  Modeling of Performance, Power, and Resilience on Petascale Systems},'' in
  [{\em The Journal of Supercomputing}{\nolinebreak\hspace{0.1em}]},  (2018).

\bibitem{Fan4}
Fan, Y. and Lan, Z., ``{Exploiting Multi-Resource Scheduling for HPC},'' in
  [{\em SC Poster}{\nolinebreak\hspace{0.1em}]},  (2019).

\bibitem{Qiao2}
Qiao, P., Wang, X., Yang, X., Fan, Y., and Lan, Z., ``{Joint Effects of
  Application Communication Pattern, Job Placement and Network Routing on
  Fat-Tree Systems},'' in [{\em ICPP Workshops}{\nolinebreak\hspace{0.1em}]},
  (2018).

\bibitem{Fan5}
Fan, Y., Lan, Z., Childers, T., Rich, P., Allcock, W., and Papka, M., ``{Deep
  Reinforcement Agent for Scheduling in HPC},'' in [{\em
  IPDPS}{\nolinebreak\hspace{0.1em}]},  (2021).

\bibitem{Fan6}
Fan, Y. and Lan, Z., ``{DRAS-CQSim: A Reinforcement Learning based Framework
  for HPC Cluster Scheduling},'' in [{\em Software
  Impacts}{\nolinebreak\hspace{0.1em}]},  (2021).

\bibitem{Fan7}
Fan, Y., Rich, P., Allcock, W., Papka, M., and Lan, Z., ``{ROME: A
  Multi-Resource Job Scheduling Framework for Exascale HPC System},'' in [{\em
  IPDPS poster}{\nolinebreak\hspace{0.1em}]},  (2018).

\end{thebibliography}
\bibliographystyle{spiebib} 

\end{document}